\documentclass[runningheads]{llncs}

\usepackage{graphicx} 
\usepackage{todonotes}
\usepackage{enumitem}
\usepackage{amsmath}
\usepackage{soul}
\usepackage{url}
\usepackage{circledsteps}
\usepackage{booktabs}
\usepackage{cite}
\usepackage{multirow}
\usepackage[T1]{fontenc}
\usepackage[multiple]{footmisc}

\usepackage[many]{tcolorbox}
\definecolor{main}{HTML}{CFCFCF}  
\definecolor{sub}{HTML}{CFCFCF}   

\tcbset{
    sharp corners,           
    colback = white,         
    before skip = 0.2cm,     
    after skip = 0.5cm       
}

\newtcolorbox{boxC}{
    colback = sub,  
    boxrule = 0pt   
}

\begin{document}

\title{Centrality Change Proneness: an Early Indicator of Microservice Architectural Degradation}

\author{Alexander Bakhtin\orcidID{0000-0003-3513-7253}\and Matteo Esposito\orcidID{0000-0002-8451-3668}\and 
 Valentina Lenarduzzi\orcidID{0000-0003-0511-5133}\and Davide Taibi\orcidID{0000-0002-3210-3990}}
\authorrunning{Bakhtin et al.}
\titlerunning{Centrality Change Proneness: an Early Indicator of MAD}

\institute{ University of Oulu, Oulu, Finland \\  \email{{firstname.lastname}@oulu.fi} }

\maketitle

\begin{abstract}

Over the past decade, the wide adoption of Microservice Architecture has required the identification of various patterns and anti-patterns to prevent Microservice Architectural Degradation. Frequently, the systems are modelled as a network of connected services.
Recently, the study of temporal networks has emerged as a way to describe and analyze evolving networks.
Previous research has explored how software metrics such as size, complexity, and quality are related to microservice centrality in the architectural network. This study investigates whether \emph{temporal} centrality metrics 
can provide insight into the early detection of architectural degradation by correlating or affecting software metrics.
We reconstructed the architecture of 7 releases of an OSS microservice project with 42 services.
For every service in every release, we computed the software and centrality metrics.  From one of the latter, 
 we derived a new metric, Centrality Change Proneness. We then explored the correlation between the metrics.
We identified 7 size and 5 complexity metrics that have a consistent correlation with centrality, while Centrality Change Proneness did not affect the software metrics, thus providing yet another perspective and an early indicator of microservice architectural degradation.


\end{abstract}

\section{Introduction}
Over the past decade, the adoption of Microservice Architecture (MSA) has led to the identification of various architectural patterns and anti-patterns to prevent Microservice Architectural Degradation (MAD)~\cite{pigazzini2020towards,bakhtin2022survey}. Detecting such anti-patterns often involves modelling the microservice system (MSS) as a network of connected services~\cite{al2022using}. 
Recently, Bakhtin et al.~\cite{bakhtin2025network} proposed to consider network centrality as a new perspective on MSA since they found the centrality metrics to not be correlated with Software Metrics (SMs) such as size, complexity, and quality. However, they only analyzed one most recent version of the studied systems and MSA networks, without considering the data \emph{temporally}.

In the field of network science, the study of temporal networks (TNs) has gained traction as a powerful way to model and analyze networks that evolve~\cite{taylor}. Such approaches can extend traditional centrality metrics into the temporal dimension, resulting in \textit{temporal} centrality metrics (TCs,~\cite{taylor, yin, liu, huang}).
Such a temporal perspective has also sparked interest in software engineering research. For instance, Saarimäki et al. \cite{saarimaki2022towards} highlight the importance of incorporating time as a critical yet underexplored dimension in software analysis.

Therefore, leveraging the promising results obtained by Bakhtin et al.~\cite{bakhtin2025network}, we extended their study by reconstructing the architecture and calculating SMs and TCs for seven releases of \textit{train-ticket} OSS MSS benchmark. In this work, we studied the correlation of TCs and SMs temporally, as well as the ability of TC to affect SMs. Our results provided the following insights:
\begin{itemize}
\item we identified 7 size and 5 complexity metrics that have a consistent and strong correlation with TCs,
\item we defined the Centrality Change Proneness (CCP) based on an existing TC, and
\item CCP rank does not affect the SMs and thus offers a completely new perspective on the MSA and an early indicator for MAD.
\end{itemize}

Our results offer an example of using temporal centrality to track architectural trends in MSA. For practitioners, we introduce the CCP score as a means to identify potentially problematic services early.
For researchers, our findings suggest exploring the relationship between code complexity and MSA, as well as investigating how centrality metrics can be leveraged to detect architectural anti-patterns such as Nano, Hublike, and Mega services. 

\textbf{Paper structure:} Section \ref{sec:related} summarizes the related works, Section \ref{sec:design} presents the Study Design, Section \ref{sec:results} presents the results of the emprical study and Section \ref{sec:discussion} provides a discussion of the outcomes, while Section \ref{sec:threats} address the Threats to Validity of the study and Section \ref{sec:conclusion} concludes the work. Additionally, the Online Appendix \cite{appendix} contains the extended theoretical background of this work.
\section{Related Work}
\label{sec:related}

Recent research endeavours have applied networks for the assessment of MSA and detection of MAD. For instance, Bakhtin et al. \cite{bakhtin2022survey} carried out a literature review to identify which Microservice API patterns can be detected leveraging several techniques, including call graphs of the MSS.
al Maruf et al. \cite{al2022using} used traces from Istio\footnote{\url{https://istio.io/}} service mesh to reconstruct the architecture of the \textit{train-ticket} MSS and perform pattern and anti-pattern detection. Brandon et al. \cite{BRANDON2020} also aggregated microservice communication obtained through dynamic analysis into static networks and used network similarity measures to group the networks into correct and anomalous ones to perform root cause analysis.
Chakraborty et al. \cite{chakraborty2023} proposed to use instance-level monitoring data to build a causal graph of service monitoring metrics enriched by information about the system architecture. 

Most applications of TNs to MSA currently rely on monitoring or tracing data obtained through dynamic analysis. For example, Angelo and Orazio \cite{angelo2024} proposed a Kubernetes scheduler that takes into account among other things the dynamic TN of microservice communications to improve cluster allocation.
Xu et al. \cite{xu2024} built a neural network that uses a TN of microservice traces as input to forecast system states. Chen et al. \cite{CHEN2023} defined a neural network with similar input to perform anomaly detection. Sun et al. \cite{sun2025} 
leveraged graph autoencoders to extract the architectural information from the TN of traces to perform root cause analysis.
To our knowledge, this work is the first attempt to apply TNs to statically reconstructed microservice networks in the context of MAD.

Abufouda and Abukwaik \cite{abufouda2017using} argued for properly adopting temporal network methods in Empirical Software Engineering. However, their review focused only on applications of TNs for developer collaboration networks as in, e.g., the work by Bakhtin et al. \cite{bakhtin2024temporal}.
\section{Empirical Study Design}
\label{sec:design}
We designed and conducted this empirical study following the guidelines by Wohlin et al.~\cite{wohlin_experimentation_2024}. 

\subsection{Goal, Research Questions, and Hypotheses}

The \textbf{goal} of our study is to analyze how the \textbf{temporal centrality} of microservices \textit{correlates} with software metrics such as size, complexity, and quality, and whether the \textbf{derived CCP score} \textit{can affect} those metrics.

Therefore, we defined two \textbf{Research Questions} (\textbf{RQs}): 

\begin{boxC}
\textbf{RQ$_1$} Does microservice temporal centrality correlate with size, complexity, or quality metrics?
\end{boxC}
Since architectural degradation is, by definition, a temporal phenomenon, we are interested to see if the presence or lack of correlation is observed throughout the development process in the different releases of the system and describe the related trends.
Previously, Bakhtin et al.~\cite{bakhtin2025network} analyzed the correlation of size, complexity, and quality metrics with network centrality metrics across 24 MSS projects. They found that there is seldom a correlation between software and centrality metrics. 
However, the authors considered only one network snapshot per system and did not look into the temporal evolution of centrality. Thus, we aim to leverage \emph{temporal} centrality metrics designed specifically for evolving networks \cite{taylor, yin, liu, huang}. Formally, to answer this research question, we conjecture and test the following \textbf{null} and \textbf{alternative} hypotheses:

\begin{itemize}[labelindent=1.5em,leftmargin=3em]
\item[\emph{\textbf{H$_{01}$}}] \textit{There is no correlation between microservice temporal centrality and size metrics.}
\item[\emph{\textbf{H$_{02}$}}] \textit{There is no correlation between microservice temporal centrality and its code complexity.}
\item[\emph{\textbf{H$_{03}$}}] \textit{There is no correlation between microservice temporal centrality and quality metrics.}
\item[\emph{\textbf{H$_{11}$}}] \textit{Microservice temporal centrality and size metrics are correlated.}
\item[\emph{\textbf{H$_{12}$}}] \textit{Microservice temporal centrality and code complexity are correlated.}
\item[\emph{\textbf{H$_{13}$}}] \textit{Microservice temporal centrality and quality metrics are correlated.}

\end{itemize}

Apart from providing time-wise centrality metrics, one novel aspect that only temporal centrality analysis can provide is a score that indicates how likely a node is to vary its centrality in the temporal network based on historical observations. Taylor et al. \cite{taylor} defined it as the First-Order Mover score (FOM).

We propose the \textbf{Centrality Change Proneness (CCP)} rank, which is obtained from the FOM score by assigning to each quartile of the FOM score a value on the ordinal scale \texttt{LOW}, \texttt{MEDIUM-LOW}, \texttt{MEDIUM-HIGH}, \texttt{HIGH}.
To analyze this newly proposed rank in the context of MAD, we formulate:

\begin{boxC}
\textbf{RQ$_2$} Can the microservice Centrality Change Proneness affect size, complexity, or quality metrics?
\end{boxC}

Taylor et al. \cite{taylor}, who defined the FOM score, proposed to use it to rank the nodes in a temporal network. 
We hypothesize that observing the CCP rank of microservices can help identify negative (or positive) trends in the service's architecture, composition, or quality that contribute to MAD.
Taylor et al. \cite{taylor} analyzed TNs of university exchange in mathematics. The authors observed that Georgia Tech University had a high FOM score, and indeed, its centrality increased from 1965 to 1975. Using knowledge about the historical period represented by the network, the authors note that this university is known to have invested heavily in mathematical research during that period.

Formally, to assess the capabilities of CCP to affect the variance of the size, complexity, and quality metrics, we pose the following \textbf{null} and \textbf{alternative} hypotheses:

\begin{itemize}[labelindent=1.5em,leftmargin=3em]
\item[\emph{\textbf{H$_{04}$}}] \textit{CCP of a microservice does not affect size metrics.}
\item[\emph{\textbf{H$_{05}$}}] \textit{CCP of a microservice does not affect complexity metrics.}
\item[\emph{\textbf{H$_{06}$}}] \textit{CCP of a microservice does not affect quality metrics.}
\item[\emph{\textbf{H$_{07}$}}] \textit{FOM score of a microservice does not correlate with the other temporal centrality metrics.}
\item[\emph{\textbf{H$_{14}$}}] \textit{CCP affects the  size metrics with statistical significance.}
\item[\emph{\textbf{H$_{15}$}}] \textit{CCP affects complexity metrics with statistical significance.}
\item[\emph{\textbf{H$_{16}$}}] \textit{CCP affects quality metrics with statistical significance.}
\item[\emph{\textbf{H$_{17}$}}] \textit{FOM score of a microservice is correlated with the other temporal centrality metrics.}
\end{itemize}

\subsection{Project Selection and Data Collection}
\label{sec:collection}
In this section, we adopt the procedures and replication package from Bakhtin et al. \cite{bakhtin2025network} for data collection.

\noindent\textbf{(1) Project selection.}
Bakhtin et al. \cite{bakhtin2025network} recently compiled a dataset of OSS MSS projects suitable for network analysis by reconstructing their architectures. From 125 GitHub Java Spring repositories, only 24 yielded reasonably complete architectures using the \textit{Code2DFD} tool \cite{Schneider23_code2dfd}. We focus on these 24 projects and examine their version and release history to identify candidates for temporal architectural analysis. Based on commit history, number of services, and discovered connections,
we noticed that the \textit{train-ticket} project \cite{zhou2018benchmarking}\footnote{\url{https://github.com/FudanSELab/train-ticket/}}  shows significant structural evolution \cite{bakhtin2024challenges}. From release \texttt{v0.0.1} to \texttt{v0.0.2}, developers removed the Register, Login, and SSO services, replacing them with User and Auth services. Between \texttt{v0.2.0} and \texttt{v1.0.0}, food-related services were restructured. This level of architectural change made \textit{train-ticket} a suitable subject for temporal analysis. According to changelogs of other MSS from the dataset, updates were limited to dependencies or libraries and not architecture. Observing the architectural networks of these projects in different releases, we noticed that they are identical and thus unsuitable for temporal network analysis.
Therefore, we selected \textit{train-ticket} and analyzed its seven published releases.

\noindent\textbf{(2) Reconstructing the Architecture and Temporal Network.} Using the exact version of the \textit{Code2DFD} tool \cite{Schneider23_code2dfd} as in \cite{bakhtin2025network}, we reconstructed and examined the architecture of each published release of \textit{train-ticket}
and merged them into a single TN capturing microservice interactions over time. Consistent with \cite{bakhtin2025network}, we applied the same preprocessing steps:

\noindent\textit{(2.1) Retain only the largest weakly connected component.} Due to false negatives in connection detection, reconstructed networks may contain unconnected components. We keep only the largest weakly connected component \footnote{\url{https://mathworld.wolfram.com/WeaklyConnectedDigraph.html}}.

\noindent\textit{(2.2) Remove databases connected to a single service.} Database nodes (e.g., \texttt{mysql}, \texttt{mongo}) represent infrastructure, not architecture, and typically have low centrality. Since their code is not part of the repository, we exclude any such database node connected to only one service, filtering out elements that follow the \emph{one database per microservice} pattern\footnote{\url{https://www.baeldung.com/cs/microservices-db-design}}\footnote{\url{https://microservices.io/patterns/data/database-per-service.html}} and avoid skewing the analysis.

\noindent\textbf{(3) Computing Temporal Centrality.} To consider the centrality in the temporal perspective, we selected four algorithms (more details in the Online Appendix\cite{appendix}) as follows: 
Taylor et al.~\cite{taylor} (Taylor algorithm) extended the eigenvector centrality measure of a static network into the temporal domain by constructing a \textit{supra-centrality matrix} of a temporal network by utilizing \textit{centrality matrices} of the network snapshots and \textit{inter-layer coupling} of nodes across snapshots;
Yin et al.~\cite{yin} (Yin algorithm) modified the Taylor algorithm~\cite{taylor} by considering snapshot similarity as the inter-layer coupling measure for nodes;
Liu et al.~\cite{liu} (Liu algorithm) follow a similar approach as Yin et al.~\cite{yin}, but they also used the sum of node degrees to construct centrality matrices;
Huang et al.~\cite{huang} (Huang algorithm) modified the Taylor algorithm~\cite{taylor} by fitting an ARMA model (Auto-regressive Moving Average, \cite{hannan1982recursive}) to sequences of nodes' degrees to determine inter-layer coupling.

We implemented these algorithms in Python using the \textit{numpy} library. Each algorithm produces several centrality metrics offering different perspectives: Joint, Conditional, Marginal Layer, and Marginal Node centralities. In addition, the Taylor algorithm provides the Time-Averaged Centrality and First-Order Mover (FOM) score, resulting in a total of 18 centrality metrics.
Since the FOM score is computed over the entire temporal network (TN) and does not yield a time series, we extended the analysis by recalculating the FOM score and the corresponding CCP rank for each release. We accumulated the TN until each release, simulating how these metrics would evolve if the system were continuously analyzed throughout development.

\noindent\textbf{(4) Computing Size, Complexity, and Quality Metrics.} 
We retrieved and stored the source code of all seven \textit{train-ticket} releases and run \textit{Understand}, \textit{JaSoMe}, and \textit{SonarQube} tools on each version.
Such tools produce metrics at the package level, so we manually mapped each package to its corresponding microservice based on its location in the project’s directory structure \cite{bakhtin2025network}. This process was repeated for each release to account for structural changes like renaming and refactoring.
Since microservices can span multiple packages, we aggregated package-level metrics into service-level ones using \texttt{Sum}, \texttt{Avg}, and \texttt{Max} as appropriate—ratios used only \texttt{Avg} and \texttt{Max}, while counts used all three.
After filtering out incomplete data, our final dataset included 25 \textit{Understand}, 55 \textit{JaSoMe}, 19 \textit{SonarQube}, and 18 temporal centrality metrics, computed for 43 microservices across 7 \textit{train-ticket} releases—forming the basis for our analysis.

\subsection{Data Analysis}
\label{sec:analysis}

In this section, we use the computed temporal centrality metrics from the architectural TN reconstructed with \textit{Code2DFD},  size metrics from \textit{Understand} and \textit{JaSoMe}, complexity metrics from \textit{JaSoMe} and \textit{SonarQube}, and quality metrics from \textit{SonarQube} tools for our analysis.

As described in Section~\ref{sec:collection}, we aggregated package-level metrics to the service level using \texttt{Sum} (87 metrics), \texttt{Avg} (60), and \texttt{Max} (47),
bringing the total number of unique software metrics to 196.

Temporal centralities were derived using only three out of four algorithms (Taylor, Liu, and Yin). We excluded the Huang algorithm due to its reliance on time series modelling across snapshots, which was unreliable given only seven snapshots. They are omitted from our analysis, but for completeness, we include them in our replication package \cite{appendix}.

\begin{figure}[t]
    \centering
    \includegraphics[width=\linewidth]{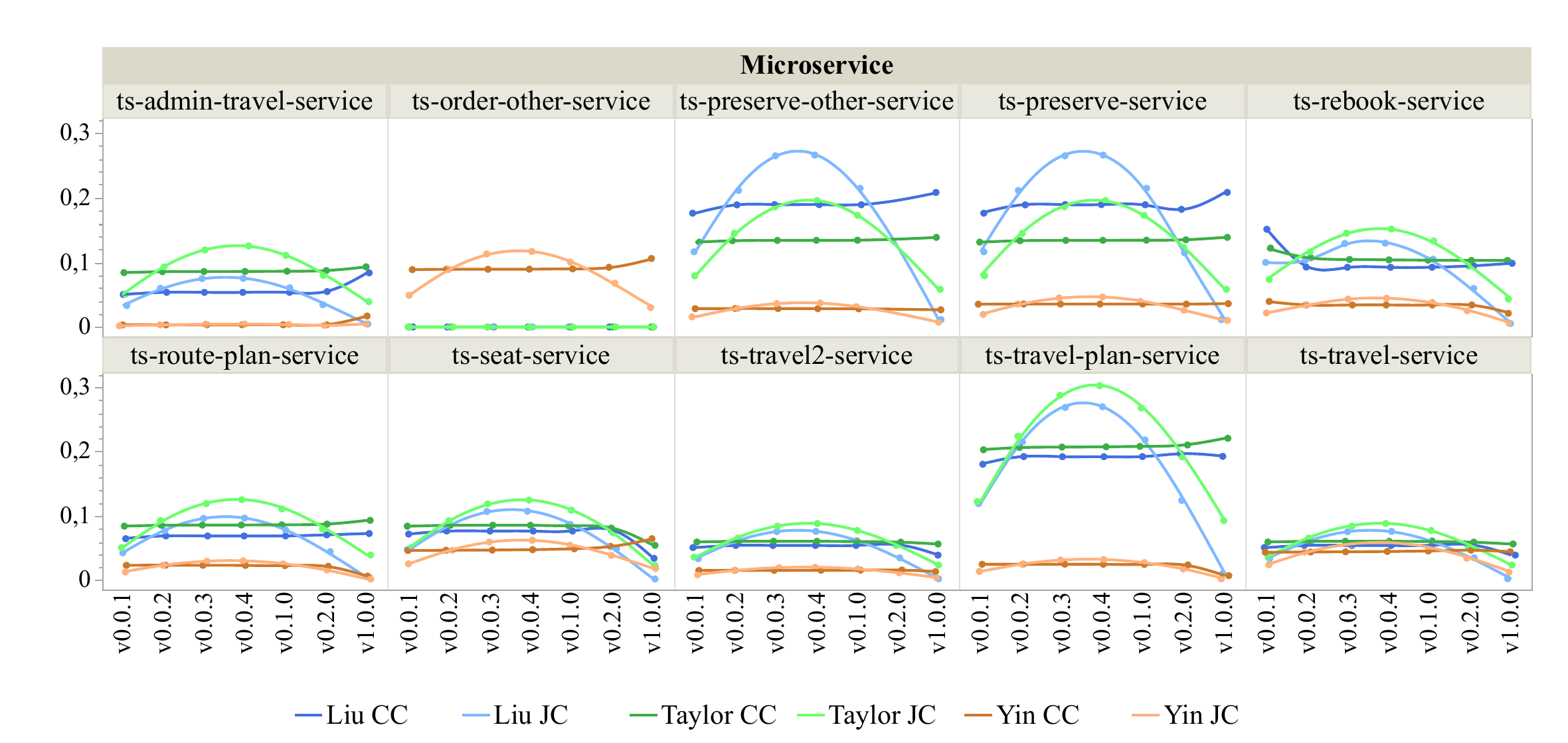}
    \caption{Microservice Joint and Conditional centrality \cite{taylor} trajectories as given by the three considered methods \cite{taylor, yin, liu} across the releases}
    \label{fig:centrality}
\end{figure}

\begin{figure} [t]
    \centering
    \includegraphics[width=\linewidth]{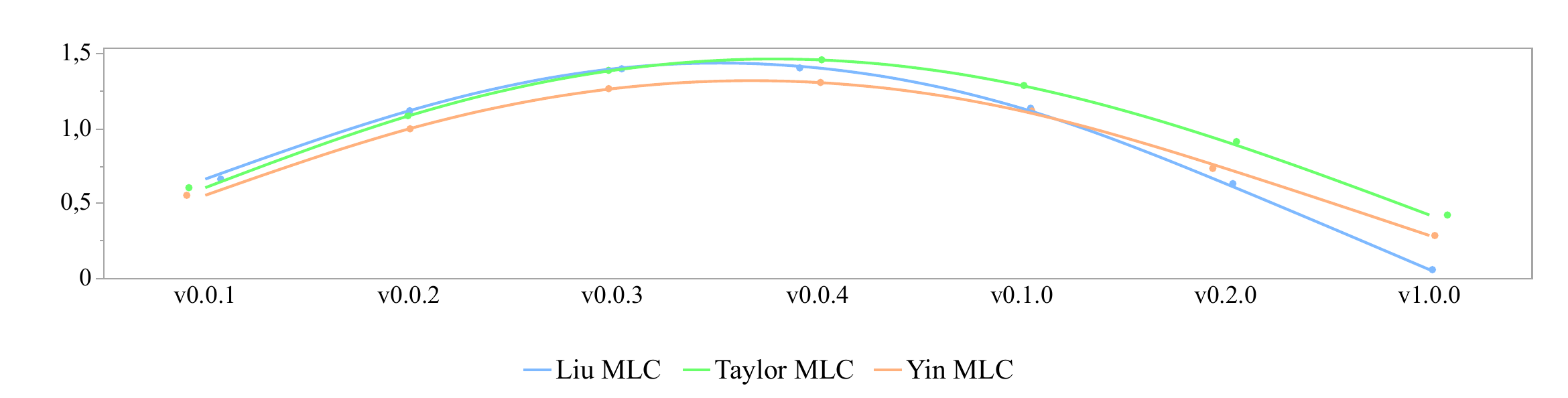}
    \caption{Sum of the centralities of all microservices (MLC, \cite{taylor}) for each release of \textit{train-ticket}}
    \label{fig:mlc}
\end{figure}

Only a subset of services exhibit significant non-zero centrality metrics (Figure \ref{fig:centrality}). This is likely due to the small size of the network and the sparsity of its adjacency matrices. Notably, the services with non-zero values are indeed the more connected and central ones. We expect that larger industrial systems would mitigate these limitations. Additionally, as noted in Section~\ref{sec:collection}, some services were introduced or removed partway through development, affecting their presence in the centrality analysis. Regarding MLC, all three algorithms resulted in a peak at release \texttt{v0.0.4} (Figure~\ref{fig:mlc}), suggesting that services in this snapshot were collectively more central than in other releases.

Finally, to select the proper statistical test, we must assess the metrics' distribution \cite{falessi_enhancing_2023}. To test the normality of the distribution of the data, we posed the following \textbf{null} and \textbf{alternative} hypotheses:

\begin{itemize}[labelindent=1.5em,leftmargin=3em]
\item[\emph{\textbf{H$_{0\mathcal{N}}$}}] The gathered metrics are normally distributed.
\item[\emph{\textbf{H$_{1\mathcal{N}}$}}] The gathered metrics are \emph{not} normally distributed.
\end{itemize}

We tested \emph{\textbf{H$_{0\mathcal{N}}$}} using Anderson-Darling (AD) test \cite{anderson1952asymptotic}.
AD tests whether data points are sampled from a specific probability distribution by
evaluating the differences between the cumulative observed distribution and the hypothesized distribution, i.e., in this case, the normal distribution.
Since we gathered metrics for seven releases of \textit{train-ticket}, we tested the hypotheses on each metric for each release separately, resulting in $1072$ hypothesis tests. We rejected 919 null hypotheses (Table \ref{tab:AD}), thus asserting that most of the gathered \textbf{metrics are not normally distributed}.

We excluded a metric if we could not reject the null hypothesis in at least one release.
This results in 50 metrics being \textbf{excluded} from further analysis.

\begin{table}[]
    \centering
\scriptsize
    \caption{Counts of rejected null hypotheses for Anderson-Darling Test}
    \label{tab:AD}
    \begin{tabular}{p{0.4cm}|l|ccccccc|r}
        &&\multicolumn{7}{c|}{Rejected null hypotheses}&Total\\ \hline
        &Version& \texttt{v0.0.1} & \texttt{v0.0.2} & \texttt{v0.0.3} & \texttt{v0.0.4} & \texttt{v0.1.0} & \texttt{v0.2.0} & \texttt{v1.0.0} &  \\
        \hline
      \multirow{3}{*}{\rotatebox{90}{Metric}}  &Size & 55 & 57 & 53 & 54 & 55 & 57 & 61 & 392 \\
       \multirow{3}{*}{} &Complexity & 46 & 48 & 47 & 49 & 50 & 54 & 51 & 345 \\
        &Quality & 25 & 27 & 25 & 28 & 28 & 27 & 22 & 182\\
        \hline
        &Total & 126 & 132 & 125 & 131 & 133 & 138 & 134 & 919 \\
        \hline
    \end{tabular}
\end{table}

To address \textbf{RQ$_1$}, since the data do not follow a normal distribution, we choose Spearman's $\rho$ \cite{spearman_rho} to test the correlation of different SMs with the TCs, instead of Pearson's test, which assumes normally distributed data\cite{pearson1895vii}. The non-parametric test evaluates the monotonic relationship between variables, assessing whether a change in one variable leads to another, either in the same direction (positive correlation) or a different (negative correlation). To interpret $\rho$ values, we adopt Dancey and Reidy's interpretation \cite{dancey2007statistics}.
We computed the correlation of each SM with each TC in each release, which resulted in $11242$ correlation tests.
Finally, we adopted a stricter significance threshold of  $\alpha = 0.01$  to balance between controlling Type I and Type II errors and maintaining statistical power while testing 11242 correlations. Due to the volume of pairwise correlation tests, the Bonferroni correction would result in a stricter threshold ($\alpha^{\prime} = \frac{0.05}{11242} = 4.4\cdot 10^{-6}$), deemed overly conservative for an exploratory study 
\cite{abac.12172}
leading to a substantial increase in the likelihood of Type II errors, i.e., potentially obscuring weaker but meaningful correlations.

To address \textbf{RQ$_2$},  we converted the normalized FOM (Norm(FOM)) score computed for each release to CCP rank by considering the quartiles of the Norm(FOM) score, and we assigned to each microservice a CCP on the ordinal scale \texttt{LOW}, \texttt{MEDIUM-LOW}, \texttt{MEDIUM-HIGH}, \texttt{HIGH} based on the corresponding quartile. This provides us with a way to rank nodes that do not depend on a predetermined threshold but on the quartiles of the FOM score specific to each studied system. In our case, the quartiles resulted in values $0.054365,0.098269,0.156734$,
We applied the Wilcoxon signed-rank test (WT, \cite{wilcoxon1945}), a non-parametric statistical method designed to compare two related samples of paired data, to test the hypotheses \emph{\textbf{H$_{14}$}}-\emph{\textbf{H$_{16}$}}.
The WT  assigns signs based on the direction of the differences between paired observations and then sums the positive and negative ranks separately. The test statistic is defined as the lowest of the two sums.
Similar to Spearman $\rho$, we use the Wilcoxon test because it does not assume a normal data distribution.
In our case, the paired data are the CCP rank and the SMs. Since we tested each SM in each release, we performed $1022$ hypothesis tests.
To test \emph{\textbf{H$_{17}$}}, we also computed the Spearman $\rho$ correlation of Norm(FOM) with all the other computed temporal centrality metrics.

Finally, we aimed to perform a time-series analysis of temporal centrality and software metrics. However, due to the small number of releases \textit{train-ticket} has, all applicable time-series methods either failed to converge or converged to null values. We believe such an analysis would be relevant when considering a bigger system with a more extended development history and is one potential direction for future work stemming from the proposed temporal centrality approach. 

\section{Results}
\label{sec:results}
In this Section, we report the results of the empirical study.

\subsection{Correlation of Centrality with Software Metrics (RQ$_1$)}

We computed the Spearman $\rho$ value for each SM and TC for each release, which provided $11242$ hypotheses. We could reject the null hypothesis only in $2434$ cases (Table \ref{tab:rq1}).
Hence, for the reminder analysis and discussion, we include only those metrics that show a \textbf{statistically significant correlation} with \textbf{at least one} centrality metric across \textbf{all releases} (Section \ref{sec:analysis}),
resulting in the inclusion of only 7 size and 5 complexity metrics, while all quality metrics were excluded.

\begin{table}[h]
    \centering
\scriptsize
    \caption{Counts of rejected null hypotheses for Spearman Rho correlation}
    \label{tab:rq1}
  \begin{tabular}{p{0.4cm}|l|ccccccc|r}
        &&\multicolumn{7}{c|}{Rejected null hypotheses}&Total\\ \hline
        &Version& \texttt{v0.0.1} & \texttt{v0.0.2} & \texttt{v0.0.3} & \texttt{v0.0.4} & \texttt{v0.1.0} & \texttt{v0.2.0} & \texttt{v1.0.0} &  \\
        \hline
  \multirow{3}{*}{\rotatebox{90}{Metric}}       &Size & 258 & 272 & 267 & 165 & 158 & 114 & 74 & 1308 \\
    \multirow{3}{*}{}       &Complexity & 237 & 216 & 188 & 78 & 78 & 50 & 33 & 880 \\
        &Quality & 75 & 42 & 21 & 35 & 29 & 23 & 21 & 246 \\
        \hline
        &Total & 570 & 530 & 476 & 278 & 265 & 187 & 128 & 2434 \\
        \hline
    \end{tabular}
\end{table}

\begin{figure}[ht]
    \centering
    \includegraphics[width=\linewidth]{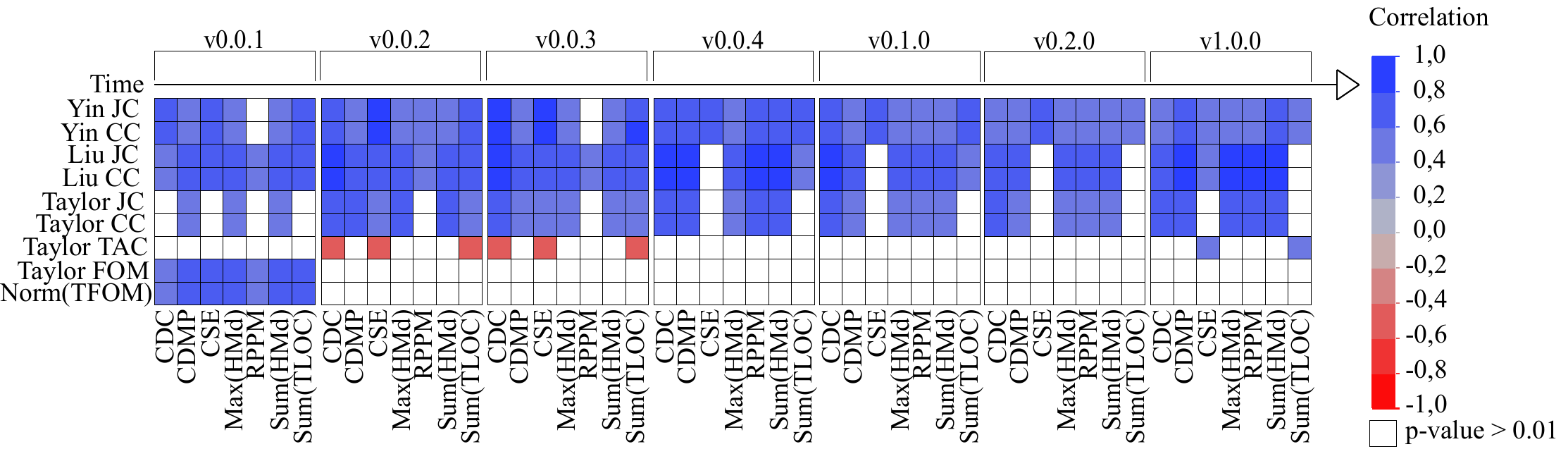}
    \caption{Heat-map of correlation of microservice temporal centrality with size metrics across the releases}
    \label{fig:correlationSize}
\end{figure}

As for the \textbf{size} metrics, we tested a total of 5,159 hypotheses and were able to reject only 1,308 (Table~\ref{tab:rq1}). Applying our metric inclusion criterion, only 7 size metrics remained. Among these, all statistically significant correlations are positive and range from moderate to strong in strength (Figure~\ref{fig:correlationSize}). For such metrics, we examined the trends in Spearman's $\rho$ values across the different releases. We see that the Ratio of Private to Public Methods (RPPM) is not correlated significantly in the first three releases, while \texttt{CountStmtExe} (CSE) and \texttt{Sum(TLOC)} are not significantly correlated in the last four releases (Figure~\ref{fig:correlationSize}). \texttt{CountDeclClass} (CDC) and \texttt{CountDeclMethodPrivate} (CDMP) are consistently positively correlated. Conversely, the first release, \texttt{v0.0.1}, differs from the others since it is the Taylor centrality metrics that fail to correlate with most software metrics rather than the other way around.
We can conclude that \textbf{size metrics} are \textbf{not correlated} with temporal centrality.

\begin{figure}[ht]
    \centering
    \includegraphics[width=\linewidth]{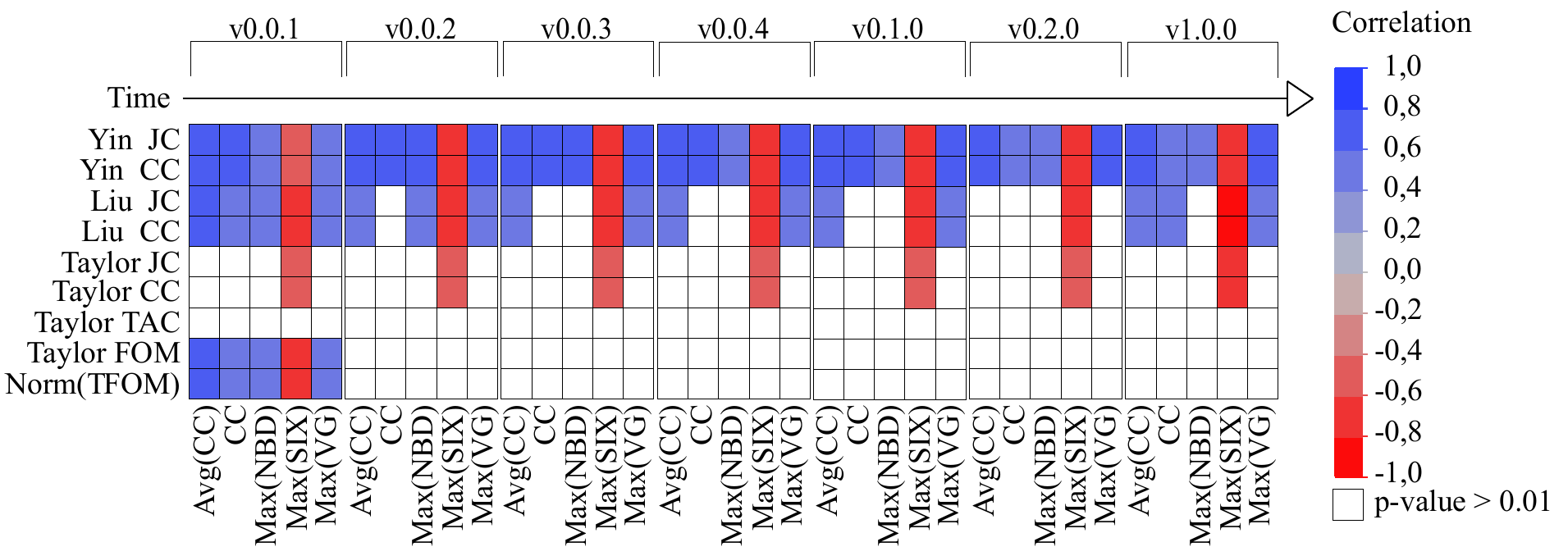}
    \caption{Heat-map of correlation of microservice temporal centrality with complexity metrics across the releases}
    \label{fig:correlationComplexity}
\end{figure}

As for the \textbf{complexity} metrics, we tested a total of 4312 hypotheses but could only reject 808 (Table \ref{tab:rq1}). We are only left with 5 metrics correlated with at least one centrality consistently (Figure \ref{fig:correlationComplexity}).
Most correlations among the selected metrics are positive, while the correlation is strong to perfectly negative for \texttt{Max(SIX)} metric. \texttt{SIX} is the \textit{specialization index} of the class from \textit{JaSoMe} tool\footnote{\url{https://github.com/rodhilton/jasome/blob/master/README.md}}, defined as $\frac{DIT*NORM}{NOM}$, where \texttt{DIT} is Depth of the Inheritance Tree, \texttt{NORM} is Number of Overridden Methods, and \texttt{NOM} is the total Number of Methods.

The average of Cognitive Complexity (Avg(CC)) seems to be consistently positively correlated in all releases, while the absolute value of Cognitive Complexity (CC) loses and gains statistical significance intermittently.

We observe that Yin metrics always correlate with all the metrics, while other centralities are mostly not correlated with anything.
Looking at the Spearman $\rho$ values trends across the releases, we see that for complexity metrics, almost the same metrics show statistically significant correlations in all releases. 
The variations in statistical significance across releases are certainly spurious.
Given that we could not reject most of the null hypotheses, we conclude that \textbf{complexity} metrics are generally \textbf{not correlated} with temporal centrality.

As for the \textbf{quality} metrics, we tested 1771 hypotheses and rejected only 246. According to our metric inclusion criterion, we could not select any metrics for further analysis and discussion since none were consistently correlated with the same centrality across the releases. We thus conclude that \textbf{quality} metrics are \textbf{not correlated} with temporal centrality.

\begin{boxC}
\emph{Overall, Microservice temporal centrality does \textbf{not} correlate with size, complexity, nor quality metrics}.
\end{boxC}

\subsection{Affect of CCP on Software Metrics (RQ$_2$)}

\begin{figure} [ht]
    \centering
    \includegraphics[width=1.0\linewidth]{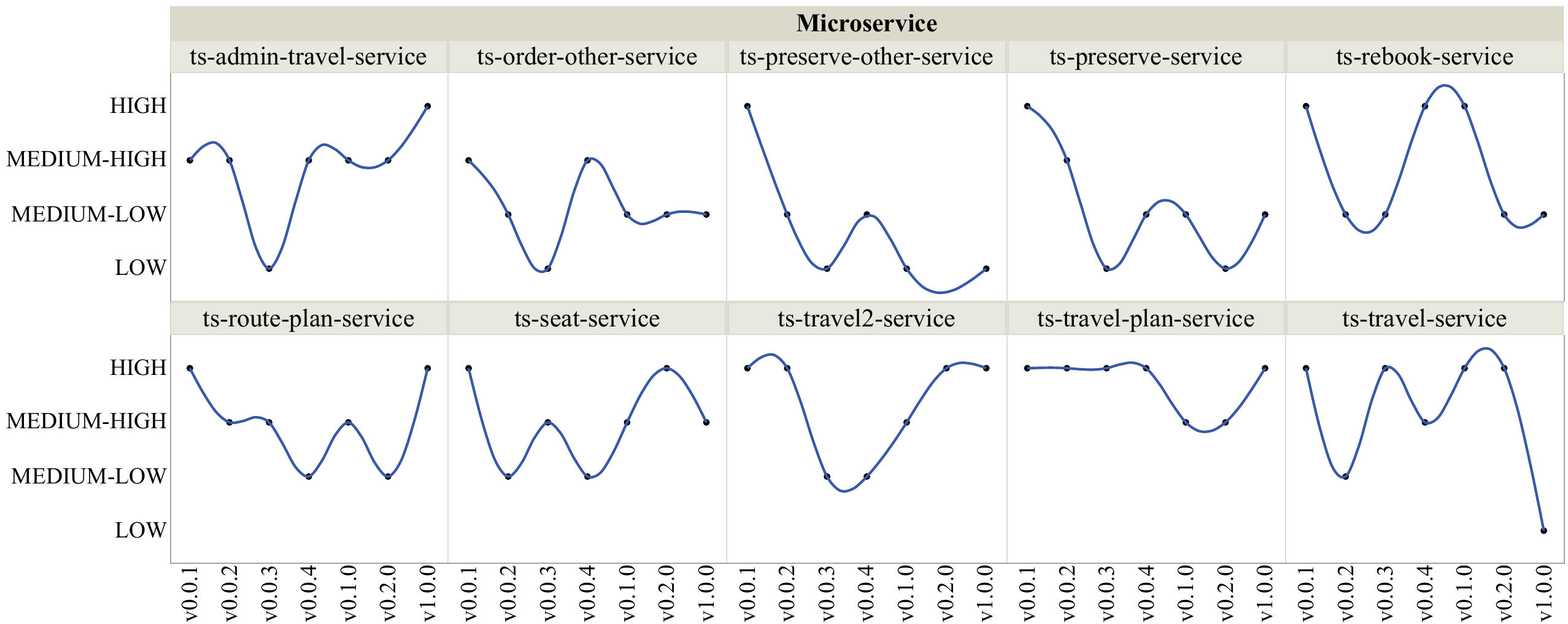}
    \caption{Trajectories of CCP rank across the releases}
    \label{fig:ccp}
\end{figure}

As described in Section \ref{sec:analysis}, we converted the FOM score to the CCP ordinal scale. Figure \ref{fig:ccp} shows the trajectories of CCP rank across releases for some microservices of \textit{train-ticket} system (same as Figure \ref{fig:centrality}).
We have tested the 1022 hypotheses of the WT and could only reject 71 (Table \ref{tab:rq2}).

We observe that most of the rejected hypotheses correspond to the first release. 
Notably, there is not a single SM that would be consistently affected by CCP across all releases. There seems to be a spurious result for one size metric at the latest, seventh release (\texttt{Max(NSF)}).
Essentially, the FOM and thus CCP scores are \emph{undefined} for the very first release since the analysis performed by Taylor et al. \cite{taylor} to derive the FOM score does not apply because then the TN consists of only one snapshot. Moreover, TAC and FOM were not correlated with SMs for most releases (RQ1, Figures \ref{fig:correlationSize}, \ref{fig:correlationComplexity}). We can thus conclude that \textbf{CCP rank} does \textbf{not affect} the variance of \textbf{size, complexity, or quality} metrics as \textit{the project matures}.
\begin{table}[h]
    \centering
\scriptsize
    \caption{Counts of rejected null hypotheses for Wilcoxon Test}
    \label{tab:rq2}
 \begin{tabular}{p{0.4cm}|l|ccccccc|r}
        &&\multicolumn{7}{c|}{Rejected null hypotheses}&Total\\ \hline
        &Version& \texttt{v0.0.1} & \texttt{v0.0.2} & \texttt{v0.0.3} & \texttt{v0.0.4} & \texttt{v0.1.0} & \texttt{v0.2.0} & \texttt{v1.0.0} &  \\
        \hline
        \multirow{3}{*}{\rotatebox{90}{Metric}}       &Size & 35 & 0 & 0 & 0 & 0 & 0 & 1 & 36 \\
      \multirow{3}{*}{}       &  Complexity & 28 & 0 & 0 & 0 & 0 & 0 & 0 & 28 \\
        &Quality & 7 & 0 & 0 & 0 & 0 & 0 & 0 & 7\\
        \hline
        &Total & 70 & 0 & 0 & 0 & 0 & 0 & 1 & 71 \\
        \hline
    \end{tabular}
\end{table}

As for the correlation of normalized FOM with centrality, we only observed a statistically significant correlation for the first release but not subsequent ones (see Online Appendix \cite{appendix}). As outlined earlier, FOM at the very first release is technically \textit{undefined}, so we can conclude that overall, the \textbf{FOM score} and, thus, \textbf{CCP rank} are \textbf{not correlated} with \textbf{temporal centrality} and thus provide \textit{a new perspective} on the evolution of the microservice architectural network.

\begin{boxC}
\emph{All in all, Microservice Centrality Change Proneness does \textbf{not} affect size, complexity, nor quality metrics.}
\end{boxC}
\section{Discussion}
\label{sec:discussion}

Observing the trends over time,  we notice a decreasing number of rejected correlation hypotheses (Table \ref{tab:rq1}).
This suggests that as the architecture matures and stabilizes, its structure and, thus, quality become increasingly decoupled from the properties of the individual microservices. After all, the architecture and its centrality are defined by an abstract graph, independent of the internal implementation of the services. \textbf{Consequently, with a stable architecture, each microservice in the MSA can vary in size, complexity, or quality based solely on how it is being developed.} This conclusion aligns with the observations of Shrikanth and Menzies~\cite{shrikanth2023} that "early bird" data, i.e., data from only a handful of the earliest commits of a project, has significant predictability for the quality of the project.
Furthermore, this corroborates the result of Bakhtin et al.~\cite{bakhtin2025network}, where the authors showed a lack of correlation between centrality and SMs for the \textit{latest} release of several MSS projects. Our temporal analysis shows that this indeed should have been the expected result.

Similarly, comparing the findings of Bakhtin et al.~\cite{bakhtin2025network} and ours, we observe a strong negative correlation between \texttt{Max(SIX)} and centrality metrics in our analysis of complexity metrics. In their study, Bakhtin et al.~\cite{bakhtin2025network} excluded the \texttt{Max(SIX)} aggregation due to its failure in the AD test, but reported that both \texttt{Avg(SIX)} and \texttt{Sum(SIX)} exhibited statistically significant strong negative correlations with degree centrality. Furthermore, they included additional complexity metrics such as \texttt{DIT} and \texttt{NORM}, which also showed moderate to strong negative correlations with degree, eigenvector, and closeness centrality scores.
This consistent pattern suggests a future research direction - \textbf{investigate the relationship between microservices’ class inheritance complexity and the quality of the resulting MSA network}, as characterized by centrality and other relevant network measures. However, since \texttt{Max(SIX)} reflects only the most complex class within a microservice, future empirical studies should consider aggregations such as \texttt{Avg} and \texttt{Sum}, which capture complexity across all constituent classes, to see if the most complex class alone or the overall complexity affect the architecture.

Furthermore, our results suggest that the proposed \textbf{CCP rank can provide a new perspective on MSA} alongside the temporal centrality trajectories, which would not require a pre-determined threshold. We see that, out of 10 services presented in Figure \ref{fig:ccp}, \textit{ts-admin-travel-service}, \textit{ts-rebook-service}, \textit{ts-route-plan-service}, \textit{ts-seat-service}, \textit{ts-travel2-service}, \textit{ts-travel-plan-service}, and \textit{ts-travel-service} have their CCP rank as \texttt{MEDIUM-HIGH} or \texttt{HIGH} for most of the releases, which is corroborated by the fact that in Figure \ref{fig:centrality}, these services are continuously changing their centrality along a U-shaped curve. We realize that for this data set, the MLC trend of peaking around release \texttt{v0.0.4} due to apparent active development at that release overshadows all other trends, so all Joint Centralities also peak at the same release and do not show other trends, while the normalized Conditional Centralities stay mostly flat, indicating a stable architecture consistent with \textit{train-ticket} being a benchmark demo app designed in advance. Release \texttt{v0.2.0} seems to be an inflection point, with centralities at the next release \texttt{v1.0.0} either rising or falling. This can be explained by the refactoring of food-related services performed while preparing that release.
Since both FOM and CCP can be computed from the early releases of a project, and given our observation that centrality becomes increasingly decorrelated from software metrics in later stages, we argue that the \textbf{CCP rank can act as an \emph{early degradation indicator} in MSA systems.} A microservice's tendency to significantly change its centrality in the network may indicate shifts in dependency structures, which could stem from flawed architectural decisions or an improper division of responsibilities. For example, as empirically validated by Palomba et al. \cite{palomba2018}, the presence of code smells can make the software components more change- and fault-prone. This insight is particularly valuable for practitioners as it offers a means to detect potentially problematic services early in the development lifecycle, enabling timely architectural adjustments.

Finally, as suggested by Bakhtin et al.~\cite{bakhtin2025network}, an important future research direction would be to explore how microservice centrality within an MSA, particularly temporal centrality, \textbf{can be leveraged to detect architectural anti-patterns} such as Nano, Hublike, or Mega services \cite{cerny2023catalog}.
\section{Threats to Validity}
\label{sec:threats}

We discuss the threats to the validity of our study, following the guidelines defined by Wohlin et al.~\cite{wohlin_experimentation_2024}.

\textbf{Construct Validity}. Our design, including our choice of tools or data filtering, may affect our results. We used existing tools Understand, JaSoMe, and SonarQube for SMs computation with no adjustment. For TCs, we created our implementation of the considered algorithms, which might be subject to human error in understanding or implementing the algorithms. The studied network is small and sparse, which could have resulted in numerical instability and provided many zero values during the computation.

\textbf{Internal Validity}.  The selection of projects to be studied can bias our results since we used a single OSS benchmark microservice project. We tried to replicate the process of Bakhtin et al. \cite{bakhtin2025network}, so we were restricted by the projects they could successfully perform the analysis on. We tried to find all suitable projects for temporal analysis with evolving MSA, but our investigation determined \textit{train-ticket} as the only suitable candidate. We hope that this study can eventually be replicated on a big industrial platform.

\textbf{External Validity}.  Mining platforms like GitHub can bias the results, since its user base comprises contributors to open-source projects, potentially skewing findings toward this specific demographic. The focus on OSS constitutes a bias since it may not represent the industrial microservice-based systems since \textit{train-ticket} is a benchmark app developed by researchers but nonetheless adopted widely in empirical studies on MSA.
Our data collection process exhibits the same threats as in \cite{bakhtin2025network} - imperfect reconstruction with the Code2DFD tool and the necessity to keep only the weak connected component and filter databases.

\textbf{Conclusion Validity}. 
Some of the used temporal centrality methods connect nodes both to the past and to the future snapshots, thus potentially causing the flow of information in a way that is not causally possible \cite{taylor, kumar}, affecting our reasoning on centrality trends in MSA. Moreover, we correlated the JC and CC values computed for the temporal network at the latest release, while FOM values were computed incrementally by release.
\section{Conclusion}
\label{sec:conclusion}
Our work further supports the notion that network centrality provides new insight into microservice architecture \cite{bakhtin2025network}. We illustrated that temporal centrality becomes less and less related to software metrics as the MSA matures, indicating that the quality of implementation and architectural structure decorrelate over time and need to be evaluated separately.

For \textbf{researchers}, our findings open some promising paths. Future research should investigate whether and how centrality, specifically temporal algorithms, can serve as a smell for MSA anti-pattern detection such as Nano, Hub-like, or Mega services. Furthermore, our findings indicate that it is necessary to better understand the effect of internal code complexity on architectural topology and stability of microservice systems.

For \textbf{practitioners}, our proposed \textit{Centrality Change Proneness} (CCP) ranking offers a lightweight, early-warning indicator of architectural volatility. Services that repetitively have high CCP rankings are most likely to undergo centrality changes, uncovering underlying responsibility misalignments or changing dependencies, i.e., a warning sign of architectural anti-patterns at an early point in time \cite{shrikanth2023, palomba2018}. 


~\\\noindent\textbf{Data Availability Statement.}
All the background, scripts, and data used for analysis
are available in our Online Appendix \cite{appendix}.

\noindent\textbf{Acknowledgment.} This work has been funded by the Research Council of Finland (grants n. 359861 and 349488 - MuFAno) and Business Finland (grant 6GSoft).

\bibliographystyle{splncs04}
\bibliography{bibliography}

\end{document}